\documentclass{MyStyle}
\usepackage{natbib}
\usepackage{hypernat}

\usepackage[notablist]{endfloat}

\author{Thierry Huillet\addressmark{1}
and Christian Paroissin\addressmark{2}}
\title[Estimating the number of species]{Sampling from Dirichlet partitions: estimating the number of species}

\address{\addressmark{1} Universit\'e de Cergy-Pontoise, Laboratoire de Physique Th\'eorique et Mod\'elisation - UMR CNRS 8089, 2 avenue Adolphe Chauvin, 95032 Cergy-Pontoise, France. Email: {\tt huillet@ptm.u-cergy.fr}  \\
\addressmark{2} Universit\'e de Pau et des Pays de l'Adour, Laboratoire de Math\'ematiques Appliqu\'ees - UMR CNRS 5142, Avenue de l'Universit\'e, 64013 Pau cedex, France. Email: {\tt cparoiss@univ-pau.fr}}



\LastVersion{\today}

\begin{document}
\maketitle

\begin{abstract}
Consider the random Dirichlet partition of the interval into $n$ fragments with parameter $\theta >0$. We recall the unordered Ewens sampling formulae from finite Dirichlet partitions. As this is a key variable for estimation purposes, focus is on the number of distinct visited species in the sampling process. These are illustrated in specific cases. We use these preliminary statistical results on frequencies distribution to address the following sampling problem: what is the estimated number of species when sampling is from Dirichlet populations? The obtained results are in accordance with the ones found in sampling theory from random proportions with Poisson-Dirichlet distribution. To conclude with, we apply the different estimators suggested to two different sets of real data.
\end{abstract}

%%%%%%%%%%%%%%%%%%%%%%%%%%%%%%%%%%%%%%%%%%%%%%%%%%%%%%%%%%%%%%%%%%%%%%%%%%%%%%
\section{Introduction}
%%%%%%%%%%%%%%%%%%%%%%%%%%%%%%%%%%%%%%%%%%%%%%%%%%%%%%%%%%%%%%%%%%%%%%%%%%%%%%%

Dirichlet partition of an interval can be viewed as a generalization of some classical models in ecological statistics. For example, on the one hand, when $\theta =1$, the Dirichlet partition corresponds to the broken-stick model (see Feller (1966), pages 22-24), one of the most famous stochastic model of relative species abundance studied by McArthur (1957) (see also Tokeshi (1993) for an exhaustive survey on species abundance models). On the other hand, when $\theta $ goes to infinity, the Dirichlet partition is deterministic and uniform and when $\theta$ goes to $0$, jointly with the numbers of fragments going to infinity, the ordered version of Dirichlet partition identifies with the Poisson-Dirichlet ($PD$) partition and corresponds to the Fisher's log-series model. These relationships between all models cited above was already pointed out early by Simpson (1949) (but the term "Dirichlet distribution" was coined by Wilks (1962) many years later).
\\[1ex]
The organization of this manuscript is the following. In Section $2$ we recall the Ewens sampling formulae when sampling is from finite Dirichlet partitions. Consider the random Dirichlet partition of the interval into $n$ fragments with parameter $\theta>0$. Elementary properties of its $D_n(\theta)$ distribution are first recalled in section $2.1$. Section $2.2$ describes some motivational sampling problems from Dirichlet proportions. Some generalities about sampling from Dirichlet partition are first proved in subsection $2.3$. Subsection $2.4$ is devoted to the Ewens sampling formulae when sampling is from Dirichlet partition $D_n(\theta)$. Here the order in which sequentially sampled species arise is irrelevant. Similarly, the second Ewens sampling formula under the same hypothesis (as a problem of random partitioning of the integers). As corollaries to these results, assuming $n\uparrow \infty $, $\theta \downarrow 0$ while $n\theta =\gamma >0$, the usual well-known sampling formulae will be deduced in each case when sampling is from $PD(\gamma)$ distribution. These general sampling formulae are also illustrated in detail in two particular cases: the Bose-Einstein case (when $\theta =1$) and the Maxwell-Bolztmann case (when $\theta $ tends to infinity). As this is the key variable for estimation purposes, focus is also made on the number of distinct visited species in the sampling process, in each case.
\\[1ex]
Section $3$ concerns the statistical problem of estimating the number of distinct species in a Dirichlet population. A maximum-likelihood estimator is developed, which is derived from sampling formulae recalled in the previous section. We recall also the minimum variance one suggested by Keener {\em et al.} (1987). For some particular classes of Dirichlet partitions, we supply simpler expressions for these estimators. We study some related statistical questions like stopping rule in the sampling process and goodness of fit. In the last subsection, we explore the difficult problem of estimating jointly $n$ and $\theta $.
\\[1ex]
At least section $4$ is devoted to applications to real data. The first two data sets concern word usage by two authors (Keener {\em et al.}, 1987) while the second two data sets deal with tropical beetles species (Janzen, 1973).

%%%%%%%%%%%%%%%%%%%%%%%%%%%%%%%%%%%%%%%%%%%%%%%%%%%%%%%%%%%%%%%%%%%%%%%%%%%%%%%
\section{Sampling from Dirichlet proportions}
%%%%%%%%%%%%%%%%%%%%%%%%%%%%%%%%%%%%%%%%%%%%%%%%%%%%%%%%%%%%%%%%%%%%%%%%%%%%%%%

First we recall basic properties about the symmetric Dirichlet distributions. Second we give some motivation about sampling with this distribution. Then we recall sampling formulae that will be useful later.

\subsection{Dirichlet partition of the interval}

Consider the following random partition into $n$ fragments of the unit interval. Let $\theta >0$ be some parameter and assume that the random fragment sizes ${\bf S}_{n}=\left( S_{1},\ldots,S_{n}\right) $ (with $\sum_{m=1}^{n}S_{m}=1$) are exchangeable and distributed according to the (symmetric) Dirichlet $D_n(\theta)$ p.d.f. which is defined on the simplex, i.e. 
\begin{equation}\label{eq1}
f_{S_1,\ldots,S_n}\left( s_1,\ldots,s_n\right) =\frac{\Gamma(n\theta)}{\Gamma( \theta)^n}\prod_{m=1}^{n}s_m^{\theta -1}\delta _{\left( \sum_{m=1}^{n}s_m-1\right)} \;.  
\end{equation}
Alternatively the distribution of ${\bf S}_{n}=\left( S_{1},\ldots,S_{n}\right) $ can also be characterized by its joint moment function 
\begin{equation*}
\forall (q_1, \ldots, q_n) \in \RR^n \;, \quad  {\bf E}_{\theta }\left[ \prod_{m=1}^{n}S_m^{q_m}\right] =\frac{\Gamma(n\theta)}{\Gamma(n\theta +\sum_{m=1}^{n}q_m)}\prod_{m=1}^{n}\frac{\Gamma (\theta +q_m)}{\Gamma(\theta)} \;. 
\end{equation*}
We shall put ${\bf S}_{n}\stackrel{d}{\sim }$ $D_{n}(\theta)$ if ${\bf S}_{n}$ is Dirichlet distributed with parameter $\theta $. In such case $S_{m}\stackrel{d}{=}S_n$, for any $m \in \{1,\ldots,n\}$, independently of $m$ and the individual fragment sizes are all identically distributed. Their common p.d.f. on the interval $(0,1) $ is the beta distribution with parameter $(\theta,(n-1)\theta)$. As a result, parameter $\theta $ interprets as a "precision" parameter indicating how concentrated the distribution of ${\bf S}_{n}$ is around its mean $(\tfrac{1}{n},\ldots,\tfrac{1}{n})$: the larger $\theta $ is, the more the distribution of ${\bf S}_{n}$ is concentrated around its mean. Indeed as one can check that, for any $m \in \{1, \ldots, n\}$, ${\bf E}(S_m)=\tfrac{1}{n}$ and ${\mbox{var}}(S_m) =\tfrac{n-1}{n^{2}(n\theta +1)}$.
\\[1ex]
In the random division of the interval as in equation (\ref{eq1}), although all fragments are identically distributed with expected sizes of order $\tfrac{1}{n}$, the smallest fragment size grows like $n^{-(\theta +1)/\theta }$ while the size of the largest is of order $\frac{1}{n\theta}\log (n\log ^{\theta -1}n)$. Consistently, the smaller $\theta $ is, the larger (resp. the smaller) the largest (resp. the smallest) fragment size is: hence, the smaller $\theta $ is, the more the values of the $S_m$ are disparate with high probability. Let ${\bf S}_{(n) }=(S_{(1)},\ldots,S_{(n)})$ be the ordered version of ${\bf S}_n$ with $S_{(1)}>\cdots>S_{(n)}$. The smaller the parameter $\theta $ is, the more the size of the largest fragment $S_{(1)}$ tends to dominate the other ones. On the contrary, for large values of $\theta$, the fragment sizes look more homogeneous and distribution equation (\ref{eq1}) concentrates on its centre $(\tfrac{1}{n},\ldots,\tfrac{1}{n})$. For large $\theta $, the diversity of the partition is small.
\\[1ex]
When $\theta =1$, the partition corresponds to the standard uniform random partition model of the interval. When $\theta \uparrow \infty $, ${\bf S}_{n}$ approaches the deterministic partition of the interval into $n$ equal parts with sizes $1/n$. Although ${\bf S}_{n}$ has a degenerate weak limit, when $n\uparrow \infty $, $\theta \downarrow 0$ while $n\theta =\gamma >0$, this situation is worth being considered. Indeed, many interesting statistical features emerge from the fact that in such asymptotic regime ${\bf S}_{\left( n\right)}$ converges to ${\bf S}_{(\infty)}$ having the Poisson-Dirichlet distribution $PD(\gamma)$ with parameter $\gamma$ (see Kingman, 1975). These three situations will be referred later respectively by Bose-Einstein, Maxwell-Boltzmann and Kingman cases.

% % % % % % % % % % % % % % % % % % % % % % % % % % % % % % % % % % % % % % %
\subsection{Sampling: motivations}
% % % % % % % % % % % % % % % % % % % % % % % % % % % % % % % % % % % % % % %

We shall be interested in sampling problems from random partition ${\bf S}_{n}$, where ${\bf S}_{n}\stackrel{d}{\sim }$ $D_{n}\left( \theta \right)$. Since ${\bf S}_{n}$ is random, sampling occurs in a random environment. Dirichlet distributions are ubiquitous in the natural sciences and this is why we chose this model for the random probabilities ${\bf S}_{n}$. We refer to Vlad {\em et al.} (2001) where it is shown that Dirichlet distributions may be seen as limit laws of certain "dilution" processes, and also that they maximize entropy under constraints, satisfy some scale-invariance property, etc. Due to its specific statistical properties as a random partition, many combinatorial issues arising in this sampling context can receive a proper and exact analytical answer. We shall illustrate this point.
\\[1ex]
Sampling from ${\bf S}_{n}$ consists in a recursive $k$ throw of iid uniformly distributed random variables on the interval. It is said that fragment number $m$ is visited by some uniform throw if its hits the interval $\left[ S_{1}+\cdots+S_{m-1},S_{1}+\cdots+S_{m}\right] $ of length $S_{m}$. Before giving some technical details of the sampling problem, let us list some motivating concrete images of the sampling problem from ${\bf S}_{n}$:

\begin{itemize}
\item $S_{m}$ could be the random abundance of species $m$ from a population with $n$ "animals". Some sampling process starts when a biologist records each new species met at each of his $k$ measurement campaigns.

\item $S_{m}$ could be the random size of district number $m$ of some city, with $m \in \{1,\ldots,n\}$ (e.g. with $n=20$ for Paris). An unfriendly sampling process could be a scattered shot bombing with $k$ bombs.

\item $S_{m}$ could be the random popularity of book $m$ in a library with $n$ books. The sampling process is when $k$ consecutive readers borrow books from the library while respecting their popularities.

\item $S_{m}$ could be the random probability to be born on day $m$, with $n=365. $ A classroom with $k$ students is a $k-$sample from ${\bf S}_{n}$.

\end{itemize}

We now give a non-exhaustive list of statistical problems of interest in this context:

\begin{itemize}

\item {\em Abundance estimation:} given a sample size $k$, estimate the number $n$ of species and/or the parameter $\theta $, exploiting for example the information on the empirical number $p$ of distinct visited species or from the knowledge of the empirical probability to visit twice the same species.

\item {\em Match box problem:} what is the state of fragment occupancies if sequential sampling process is stopped when some fragment has received $c$ visits for the first time (if $n=2$, this is the randomized Banach match box problem). In particular what is the probability that some cell is empty at this stopping time?

\item {\em Birthday problem:} what is the sample size until the first visit to two species of the same type? 

\item {\em Coupon collector problem:} what is the sample size until all species have been visited at least once (or $r$ times)? 

\item {\em Law of succession:} given the random number of occurrences of species $m$ in the $k-$sample and the number $p$ of distinct visited species, what is the probability to discover a new species in a $k+1$ sample? or what is the probability that the $(k+1)$-th sample is one from the previously encountered species already met a certain amount of times.
\end{itemize}

We shall now be more precise and treat rigorously some of the raised problems, starting with the sampling problem before focusing on the occupancy distributions. In the next section, using these results, we shall come to the important problem of estimating $n$ when it is unknown.

% % % % % % % % % % % % % % % % % % % % % % % % % % % % % % % % % % % % % % %
\subsection{Sampling: preliminaries and generalities}
% % % % % % % % % % % % % % % % % % % % % % % % % % % % % % % % % % % % % % %

Let $\left( U_{1},\ldots,U_{k}\right) $ be $k$ iid uniform throws on ${\bf S}_{n}$. Let 
\begin{equation*}
{\bf K}_{n,k}=\left( {\cal K}_{n,k}\left( 1\right) ,\ldots,{\cal K}_{n,k}\left( n\right) \right) \geqslant 0 
\end{equation*}
be an integral-valued random vector which counts the number of visits to the different fragments in a $k$-sample. Hence, if $M_{l}$ is the random fragment number (or label) in which the $l$-th trial falls, ${\cal K}_{n,k}\left( m\right) =\sum_{l=1}^{k}{\bf I}\left( M_{l}=m\right) $, $m \in \{1,\ldots,n\}$. Under our assumptions, for instance, we have that ${\bf P}_{\theta }\left( M_{l}=m\mid {\bf S}_{n}\right) =S_{m}$ (the random probability to visit species $m$ is equal to its abundance) but also that the conditional probability to observe species number $m$ in a $k$-sample is 
\begin{equation}\label{e4}
\pi _{m}\left( S_{m}\right) ={\bf P}_{\theta }\left( {\cal K}_{n,k}\left( m\right) >0\mid {\bf S}_{n}\right) =1-\left( 1-S_{m}\right) ^{k}. 
\end{equation}

Let us now focus our attention on the distribution of the occupancies ${\bf K}_{n,k}$. With $\sum_{m=1}^{n}k_{m}=k$ and ${\bf k}_{n}=\left( k_{1},\ldots,k_{n}\right)
\geqslant 0$, ${\bf K}_{n,k}$ follows the conditional multinomial distribution: 
\begin{equation*}
{\bf P}_{\theta }\left( {\bf K}_{n,k}={\bf k}_{n}\mid {\bf S}_{n}\right) = \frac{k!}{\prod_{m=1}^{n}k_{m}!}\prod_{m=1}^{n}S_{m}^{k_{m}} \;.  
\end{equation*}
Averaging over ${\bf S}_{n}$, using Dirichlet integrals, one finds 
\begin{equation*}
{\bf P}_{\theta }\left( {\bf K}_{n,k}={\bf k}_{n}\right) ={\bf EP}_{\theta}\left( {\bf K}_{n,k}={\bf k}_{n}\mid {\bf S}_{n}\right) =\frac{\prod_{m=1}^{n}\left[ \theta \right] _{k_{m}}^{n}}{\left[ n\theta \right]_{k}},  
\end{equation*}
where $[\theta]_k=(\theta)_k/k!$ and $(\theta)_k=\theta(\theta +1)\cdots(\theta+k-1)$,for any  $k\geqslant 1$, with $(\theta)_0=1$. Applying Bayes formula, the posterior distribution of ${\bf S}_n$ given ${\bf K}_{n,k}={\bf k}_n$ is determined by its p.d.f. at point ${\bf s}_n$on the simplex as 
\begin{equation*}
f_{{\bf S}_n}({\bf s}_n\mid {\bf K}_{n,k}={\bf k}_n) =\frac{\Gamma(n\theta +k)}{\prod_{m=1}^{n}\Gamma (\theta+k_m) }\prod_{m=1}^{n}s_m^{(\theta +k_m)-1} \,\delta _{(\sum_{m=1}^{n}s_m-1) }. 
\end{equation*}
This shows, as it is well-known, that 
\begin{equation*}
{\bf S}_{n}\mid {\bf K}_{n,k}={\bf k}_{n}\stackrel{d}{\sim }D_{n}\left(\theta {\bf 1}+{\bf k}_{n}\right) ,  
\end{equation*}
an asymmetric Dirichlet distribution with parameters $\theta {\bf 1}+{\bf k}_{n}=\left( \theta +k_{1},\ldots,\theta +k_{n}\right) $. Furthermore, 
\begin{equation*}
\forall m \in \{1,\ldots,n \} \;, \quad {\bf E}_{\theta }\left( S_{m}\mid {\bf K}_{n,k}={\bf k}_{n}\right) =\frac{\theta +k_{m}}{n\theta +k}  \;. 
\end{equation*}
This suggests a recursive approach to the sampling formula where successive sample are drawn from the corresponding iterative posterior distributions. More specifically, let $\left( M_{1},\ldots,M_{k}\right) \in \left\{1,\ldots,n\right\} ^{k}$ be the numbers of the successive fragments thus drawn.
Then, 
\begin{equation*}
{\bf P}_{\theta }\left( M_{1}=m_{1}\right) ={\bf E}\left( {\bf P}_{\theta}\left( M_{1}=m_{1}\right) \mid {\bf S}_{n}\right) ={\bf E}_{\theta }\left(S_{m_{1}}\right) =\frac{\theta }{n\theta }=\frac{1}{n} \;, 
\end{equation*}
\begin{equation*}
{\bf P}_{\theta }\left( M_{2}=m_{2}\mid M_{1}\right) =\frac{\theta +{\bf I}\left( M_{1}=m_{2}\right) }{n\theta +1}
\end{equation*}
and 
\begin{equation*}
{\bf P}_{\theta }\left( M_{k}=m_{k}\mid M_{1},\ldots,M_{k-1}\right) =\frac{\theta +\sum_{l=1}^{k-1}{\bf I}\left( M_{l}=m_{k}\right) }{n\theta +k-1}. 
\end{equation*}
Proceeding in this way, the joint distribution of $\left(M_{1},\ldots,M_{k}\right) $ reads 
\begin{equation*}
{\bf P}_{\theta }\left( M_{1}=m_{1},\ldots,M_{k}=m_{k}\right) = \frac{\theta }{n\theta }\prod_{l=1}^{k-1}\frac{\theta +\sum_{j=1}^{l}{\bf I}\left(M_{j}=m_{l+1}\right) }{n\theta +l}  = \frac{\prod_{m=1}^{n}\left( \theta \right) _{k_{m}}}{\left( n\theta\right){k}},
\end{equation*}
where $k_{m}=\sum_{l=1}^{k}{\bf I}\left( m_{l}=m\right)$. This distribution being invariant under permutations of the entries, the sequence $\left( M_{1},\ldots,M_{k}\right) $ is exchangeable. It is called a P\`{o}lya urn sequence. We now prove the following convergence result:

\begin{lemma}
Almost surely and in distribution, the following convergence holds: 
\begin{equation*}
\frac{{\bf K}_{n,k}}{k} \xrightarrow[k\rightarrow \infty]{a.s.} {\bf S}_{n} \;.
\end{equation*}
\end{lemma}

\noindent
{\bf Proof} Let us first prove the convergence in distribution. The joint conditional generating function of ${\bf K}_{n,k}$ reads 
\begin{equation*}
{\bf E}_{\theta }\left( \prod_{m=1}^{n}u_{m}^{{\cal K}_{n,k}\left( m\right)}\mid {\bf S}_{n}\right) =\left( \sum_{m=1}^{n}u_{m}S_{m}\right) ^{k}, 
\end{equation*}
which is homogeneous with degree $k$ allowing to compute ${\bf E}_{\theta}\left( \prod_{m=1}^{n}u_{m}^{{\cal K}_{n,k}\left( m\right) }\right)$. Further, defining $\widetilde{X}_{m}=X_{m}/\sum_{m=1}^{n}X_{m}$, where $X_{m}\stackrel{d}{\sim }$ gamma$\left( \theta \right) $, for all $m \in \{1,\ldots,n\}$, using independence between $(\widetilde{X}_1,\ldots,\widetilde{X}_n)$ and $\sum_{m=1}^{n}X_{m}$ and recalling that $(\widetilde{X}_1,\ldots,\widetilde{X}_n)$ has the Dirichlet distribution $D_n(\theta)$, we get 
\begin{eqnarray*}
{\bf E}_{\theta }\left( \prod_{m=1}^{n}u_{m}^{{\cal K}_{n,k}\left( m\right)/k}\right) 
 & = & \frac{\Gamma \left( n\theta \right) }{\Gamma \left( n\theta+k\right) }{\bf E}_{\theta }\left[ \left( \sum_{m=1}^{n}u_{m}^{1/k}X_m\right) ^{k}\right] \\ 
 & \underset{k\rightarrow \infty }{\sim } & \frac{\Gamma \left( n\theta \right) }{\Gamma \left( n\theta +k\right) }{\bf E}_{\theta }\left[ \left(\sum_{m=1}^{n}X_{m}\right) ^{k}\left( 1+\frac{1}{k}\sum_{m=1}^{n}\widetilde{X}_{m}\log u_{m}\right) ^{k}\right] \\
 & \underset{k\rightarrow \infty }{\sim } & {\bf E}_{\theta }\left( \prod_{m=1}^{n}u_{m}^{\widetilde{X}_{m}}\right) ={\bf E}_{\theta }\left( \prod_{m=1}^{n}u_{m}^{S_{m}}\right) . 
\end{eqnarray*}
Thus, 
\begin{equation*}
\frac{{\bf K}_{n,k}}{k} \xrightarrow[k\rightarrow \infty]{a.s.} {\bf S}_{n} \;.
\end{equation*}
By applying the strong law of large numbers (conditionally given ${\bf S}_{n} $), the above convergence in distribution also holds almost surely. This shows that ${\bf K}_{n,k}/k$ can be used as an consistent estimator of ${\bf S}_{n}$. $\square$

% % % % % % % % % % % % % % % % % % % % % % % % % % % % % % % % % % % % % % %
\subsection{Ewens sampling formulae for Dirichlet partitions}
% % % % % % % % % % % % % % % % % % % % % % % % % % % % % % % % % % % % % % %

Ewens Sampling Formula (ESF) gives the distribution of alleles (different types of genes) in a sample with size $k$ from the Poisson-Dirichlet partitioning $PD\left( \gamma \right) $. Alternatively, it can be described in terms of sequential sampling of animals from a countable collection of distinguishable species drawn from $PD(\gamma) $. It provides the probability of the partition of a sample of, say $k$, selectively equivalent genes into a number of alleles as population size becomes indefinitely large. When the order of appearance of sequentially sampled species does not matter, we are led to the first ESF for unordered sequences. A second equivalent way to describe the sample is to record the number of species in the $k$-sample with exactly $i$ representatives, for $i \in \{0,\ldots,k\}$. When doing this while assuming the species have random frequencies following $PD(\gamma)$ distribution, we are led to a second Ewens Sampling Formula.
\\[1ex]
We recall here the exact expressions of both first and second Ewens sampling formulae, when sampling is first from finite Dirichlet random partitions with $n$ fragments. Here, the order in which the consecutive animals are being discovered in the sampling process is irrelevant. In the sampling formulae, the joint event that there are $p$ distinct fragments visited will also be taken into account. These sampling formulae give both ESF formulae from $PD(\gamma) $ when passing to the Kingman limit.
\\[1ex]
Let ${\bf S}_{n}$ be the above Dirichlet random partition with parameter $\theta >0$. Let $k>1$ and $\left( U_{1},\ldots,U_{k}\right) $ be $k$ iid uniform random throws on $\left[ 0,1\right] $. Let then $\left(M_{1},\ldots,M_{k}\right) $ be the (conditionally iid) corresponding animals species, with common conditional and unconditional distributions:  
\begin{equation*}
\forall m\in \{1,\ldots,n\} \;, \qquad {\bf P}_{\theta }\left( M=m\mid {\bf S}_{n}\right) =S_{m} \;,
\end{equation*}
and
\begin{equation*}
\forall m\in \{1,\ldots,n\} \;, \qquad {\bf P}_{\theta }\left( M=m\right) ={\bf E}\left[ {\bf P}_{\theta }\left( M=m\mid {\bf S}_{n}\right) \right] ={\bf E}_{\theta }\left( S_{m}\right) = \frac{1}{n}\;. 
\end{equation*}
Recall ${\cal K}_{n,k}\left( m\right) =\sum_{l=1}^{k}{\bf I}\left( M_{l}=m\right) $ counts the random number of occurrences of species $m$ in the $k$-sample and let $P_{n,k}=\sum_{m=1}^{n}{\bf I}\left( {\cal K}_{n,k}\left( m\right) >0\right) $ count the number of distinct species which have been visited in the $k$-sampling process.
\\[1ex]
There are two occupancies variables of interest: the first one will lead to the first Ewens sampling formula while the second corresponds to the second Ewens sampling formula. 

\begin{itemize}
\item[${\bf I.}$] For any $q \in \{1,\ldots,p\}$, ${\cal B}_{n,k}(q)>0$ is the numbers of animals of species $q$ where the $P_{n,k}=p$ species observed were labelled in an arbitrary way (independently of the sampling mechanism). Thus ${\cal B}_{n,k}$ differs from ${\cal K}_{n,k}$ in the sense that all the components of ${\cal B}_{n,k}$ are positive.
\item[${\bf II.}$] For any $i\in \{ 0,\ldots,k\}$, ${\cal A}_{n,k}(i)$ is the number of species in the $k$-sample with $i$ representatives, i.e.
\begin{equation*}
{\cal A}_{n,k}(i) =\#\{ m\in \{ 1,\ldots,n\} : {\cal K}_{n,k}(m) =i\} =\sum_{m=1}^{n}{\bf I}\left( {\cal K}_{n,k}\left( m\right) =i\right) .  
\end{equation*}
Then $\sum_{i=0}^{k}{\cal A}_{n,k}\left( i\right) =n$ is the (unknown) number of fragments and $\sum_{i=1}^{k}{\cal A}_{n,k}\left( i\right) =p$ is the number of fragments visited by the $k$-sample and ${\cal A}_{n,k}\left( 0\right) $ the number of unvisited ones. Note that $\sum_{i=1}^{k}i{\cal A}_{n,k}\left( i\right) =k$ is the sample size. The random vector $\left( {\cal A}_{n,k}\left( 1\right) ,\ldots,{\cal A}_{n,k}\left( k\right) \right) $ is called the fragment vector count or the species vector count in biology, see Ewens (1990).
\end{itemize}

For each of the two sampling problems, we easily obtain the Ewens sampling formulae from finite partitions ${\bf S}_{n}$ drawn from Dirichlet distribution. The following result can be found in Huillet (2005) (see also Ewens (1972) for the PD case)

\begin{theorem}\label{Theo6}
\hspace{0.5cm}
\begin{itemize}
\item[${\bf I}$] For any $(b_1, \ldots, b_p)$ such that $\forall q \in \{1, \ldots, p\}$, $b_q \geqslant 1$ and $\sum_{q=1}^{p}b_{q}=k$, we have 
\begin{equation}\label{e5}
{\bf P}_{\theta }\left( {\cal B}_{n,k}\left( 1\right) = b_{1},\ldots,{\cal B}_{n,k}\left( p\right) =b_{p};P_{n,k}=p\right) = \binom{n}{p}\frac{k!}{\prod_{q=1}^{p}b_{q}!}\frac{1}{\left( n\theta \right)_{k}}\prod_{q=1}^{p}\left( \theta \right) _{b_{q}}
\end{equation}

\item[${\bf II}$] For any $(a_1, \ldots, a_k) \geqslant 0$ such that $\sum_{i=1}^{k}ia_{i}=k$ and $\sum_{i=1}^{k}a_{i}=p$, we have 
\begin{equation}\label{e12}
{\bf P}_{\theta }\left( {\cal A}_{n,k}\left( 1\right) =a_{1},\ldots,{\cal A}_{n,k}\left( k\right) =a_{k};P_{n,k}=p\right) =\frac{n!}{\left( n-p\right) !}\frac{k!}{\prod_{i=1}^{k}\left(i!^{a_{i}}a_{i}!\right) }\frac{1}{\left( n\theta \right) _{k}}\prod_{i=1}^{k}\left( \theta \right) _{i}^{a_{i}}
\end{equation}
\end{itemize}
\end{theorem}

From equation (\ref{e5}) or equivalently from equation (\ref{e12}), one can obtain the marginal distribution of $P_{n,k}$:

\begin{theorem}\label{Theo6b}
For any $p \geqslant 1$,
\begin{equation}\label{e6}
{\bf P}_{\theta }\left( P_{n,k}=p\right) =\frac{n!}{\left( n-p\right) !}\frac{1}{\left( n\theta \right) _{k}}B_{k,p}\left( \theta \right) 
\end{equation}
where 
\begin{equation*}
B_{k,p}(\theta) 
 = \frac{k!}{p!}\sum_{\substack{b_q\geqslant 1 \\ \sum_{q=1}^{p}b_q=k}}\prod_{q=1}^{p}\frac{(\theta)_{b_q}}{b_q!} 
 = \sum_{\substack{a_i\geqslant 0 \\ \sum_{i=1}^{k}a_i=p \\ \sum_{i=1}^{k}ia_i=k}}\frac{k!}{\prod_{i=1}^{k}(i!^{a_i}a_i!)}\prod_{i=1}^{k}(\theta)_i^{a_i}\;.
\end{equation*}
\end{theorem}

We recall below a straightforward representation of the probability ${\bf P}_{\theta}(P_{n,k}=p)$ under the form of an alternate sum (see for example Keener {\em et al.}, pages 1471--1472).

\begin{proposition}
For any $m \in \{0,\ldots,n-1\}$, let $\left\langle \theta \right\rangle _{n,k;m}=\frac{((n-m)\theta)_{k}}{(n\theta)_{k}}$. The distribution of $P_{n,k}$ is given by
\begin{equation}\label{as3}
{\bf P}_{\theta }\left( P_{n,k}=p\right) =\sum_{q=1}^{p}\left( -1\right)^{p-q}\binom{n}{p}\binom{p}{q}\left\langle \theta \right\rangle _{n,k;n-q}.
\end{equation}
\end{proposition}

Let us now focus on two problems related to sampling as explained previously.

\paragraph{The law of succession} We would like to briefly recall a related question raised in Donnelly (1986) and Ewens (1996), concerning the law of succession.
\begin{enumerate} 
\item Let the "$M_{k+1}$ is new" denote the event that $M_{k+1}$ is none of the previously observed species. One can prove that
\begin{equation}\label{ex1}
{\bf P}_\theta(M_{k+1}\text{is new}\mid {\cal B}_{n,k}(1) =b_1,\ldots,{\cal B}_{n,k}(p)=b_p;P_{n,k}=p) =\frac{(n-p)\theta}{n\theta +k},  
\end{equation}
which is independent of cell occupancies $b_1,\ldots,b_p$ but depends on the number $p$ of distinct species already visited by the $k$-sample. With $k=p=2$, this is the probability that the first two random throws will visit any two distinct species. The complementary probability that it does not is thus $1-\tfrac{(n-1)\theta}{n\theta +1}=\tfrac{\theta +1}{n\theta +1}$. The probability to visit any fragment twice varies between $1$ and $\tfrac{1}{n}$ when $\theta $ varies from $0$ (the largest fragment dominates) to infinity (fragment sizes distribution approaches $\tfrac{1}{n}$).
\item Similarly, let the event "$M_{k+1}$ is a species seen $b_r$ times" denote the fact that the $(k+1)$-th sample is one from the previously encountered fragment already visited $b_r$ times. We easily get 
\begin{equation}\label{ex2}
{\bf P}_\theta(M_{k+1} \text{ is a species seen }b_r\text{times}\mid {\cal B}_{n,k}(1)=b_1,\ldots,{\cal B}_{n,k}(p) =b_p;P_{n,k}=p) =\frac{\theta +b_r}{n\theta+k}  
\end{equation}
which is as previously independent cell occupancies but also of the number $p$ of distinct species. 
\end{enumerate}

\paragraph{The number of distinct observations} From equations~(\ref{ex1}) and (\ref{ex2}), we also have the transition probabilities 
\begin{equation*}
{\bf P}_\theta(P_{n,k+1}=p+1\mid P_{n,k}=p) =\frac{(n-p)\theta }{n\theta +k} 
\end{equation*}
and
\begin{equation*}
{\bf P}_\theta(P_{n,k+1}=p\mid P_{n,k}=p) =\frac{\sum_{r=1}^{p}(\theta +b_r)}{n\theta +k}=\frac{p\theta +k}{n\theta +k}. 
\end{equation*}
It follows that we have the following recursion for the distribution of $P_{n,k}$: 
\begin{equation*}
{\bf P}_\theta(P_{n,k+1}=p) =\frac{(n-p+1)\theta}{n\theta+k}{\bf P}_\theta(P_{n,k}=p-1) +\frac{p\theta +k}{n\theta +k}{\bf P}_\theta(P_{n,k}=p) \;. 
\end{equation*}
Using equation (\ref{e6}), we obtain the following triangular recurrence for the quantities $B_{k,p}(\theta)$ 
\begin{equation*}
B_{k+1,p}\left( \theta \right) =\theta B_{k,p-1}\left( \theta \right)+\left( p\theta +k\right) B_{k,p}\left( \theta \right) \;. 
\end{equation*}
These should be considered with boundary conditions 
\begin{equation*}
B_{k,0}\left( \theta \right) =B_{0,p}\left( \theta \right) =0, 
\end{equation*}
except for $B_{0,0}(\theta)=1$. Under this form, $B_{k,p}(\theta) $ turns out to be the Bell polynomial in the variables $x_1=(\theta)_1, x_2=(\theta)_2,\ldots, x_k=(\theta)_k$. This leads in particular to $B_{k,1}(\theta)=(\theta)_k$, $k\geqslant 1$ and to
\begin{equation*}
{\bf P}_{\theta}(P_{n,k}=1) =\frac{n(\theta)_k}{(n\theta)_k} \;. 
\end{equation*}

\paragraph{Special cases} Let us now study the three special cases mentioned in the introduction of this section.
\begin{enumerate} 
\item {\bf Bose-Einstein case.} When $\theta =1$, equation~(\ref{e5}) simplifies to 
\begin{equation*}
{\bf P}_1({\cal B}_{n,k}(1) =b_1,\ldots,{\cal B}_{n,k}(p) =b_p;P_{n,k}=p) =\frac{\binom{n}{p}}{\binom{n+k-1}{k}}\;,  
\end{equation*}
which is independent of the cell occupancies $(b_1,\ldots,b_p)$ (i.e. the probability is uniform). As there are $\binom{k-1}{p-1}$ sequences $b_q \geqslant 1$ for all $q \in \{1,\ldots,p\}$, satisfying $\sum b_q=k$, we get $B_{k,p}(1) =\frac{k!}{p!}\binom{k-1}{p-1}$ (called Lah numbers) and 
\begin{equation*}
\forall p \in \{1,\ldots,n\wedge k \}  \;, \quad {\bf P}_1(P_{n,k}=p) =\frac{\binom{n}{p}\binom{k-1}{p-1}}{\binom{n+k-1}{k}} \;.
\end{equation*}
Equation~(\ref{e12}) reduces to 
\begin{equation*}
{\bf P}_1({\cal A}_{n,k}(1) =a_1,\ldots,{\cal A}_{n,k}(k) =a_k;P_{n,k}=p) =\frac{p!\binom{n}{p}}{\binom{n+k-1}{k}}\frac{1}{\prod_{i=1}^{k}a_i!}  \;.
\end{equation*}

\item {\bf Maxwell-Boltzmann case.} As $\theta \uparrow \infty$, the probability displayed in equation~(\ref{e5}) converges to 
\begin{equation*}
{\bf P}_\infty({\cal B}_{n,k}(1) =b_1,\ldots,{\cal B}_{n,k}(p) =b_p;P_{n,k}=p) =\binom{n}{p}\frac{k!}{\prod_{q=1}^{p}b_q!}\frac{1}{n^k} \;.  
\end{equation*}
With $S_{k,p}$ the second kind Stirling numbers, we get 
\begin{equation}\label{e8b}
\forall p \in \{1,\ldots,k\}   \;, \quad {\bf P}_\infty(P_{n,k}=p) =\frac{n!S_{k,p}}{(n-p)!n^k} \;.
\end{equation}
This result is ancient and well-known (see Johnson and Kotz, 1969). Equation~(\ref{e12}) convergences to
\begin{equation*}
{\bf P}_\infty({\cal A}_{n,k}(1) =a_1,\ldots,{\cal A}_{n,k}(k) =a_k;P_{n,k}=p) =\frac{n!k!n^{-k}}{(n-p)!\prod_{i=1}^{k}(i^{a_i}a_i!)} \;.  
\end{equation*}

\item {\bf Kingman case.} Consider the situation where $n\uparrow \infty $, $\theta \downarrow 0$ while $n\theta =\gamma >0$. In such case, the probability displayed in equation~(\ref{e5}) converges to 
\begin{equation*}
{\bf P}_\gamma^*({\cal B}_{k}(1) =b_1,\ldots,{\cal B}_{k}(p) =b_p;P_{k}=p) =\frac{k!}{p!}\frac{\gamma^p}{(\gamma)_k\prod_{q=1}^{p}b_q} \;. 
\end{equation*}
With $s_{k,p}$ the absolute value of the first kind Stirling numbers, we get 
\begin{equation}\label{e9}
\forall p \in \{1,\ldots,k  \}  \;, \quad {\bf P}_\gamma^*(P_k=p)=\frac{\gamma^{p}s_{k,p}}{(\gamma)_k} \;.
\end{equation}
It follows that the probabilities displayed in examples (\ref{ex1}) and (\ref{ex2}) converge respectively to 
\begin{equation}\label{ex3}
\frac{\gamma }{\gamma +k} \quad {\mbox{and}} \quad \frac{b_r}{\gamma +k} \;.
\end{equation}
We note also that the distribution of $P_k$ in this case is in the class of exponential families. We recall the important result of Korwar and Hollander (1973): 
\begin{equation*}
\frac{P_{k}}{\log k} \xrightarrow[k\rightarrow \infty]{a.s.} \gamma \;.
\end{equation*}
At least, in the Kingman limit, the probability displayed in (\ref{e12}) converges to 
\begin{equation*}
{\bf P}_\gamma^*({\cal A}_{k}(1) =a_1,\ldots,{\cal A}_{k}(k) =a_k;P_k=p) =\frac{k!\gamma^p}{(\gamma)_k\prod_{i=1}^{k}(i^{a_i}a_i!)} \;.
\end{equation*}
\end{enumerate} 

%%%%%%%%%%%%%%%%%%%%%%%%%%%%%%%%%%%%%%%%%%%%%%%%%%%%%%%%%%%%%%%%%%%%%%%%%%%%%%%
\section{Estimation of the number of species}
%%%%%%%%%%%%%%%%%%%%%%%%%%%%%%%%%%%%%%%%%%%%%%%%%%%%%%%%%%%%%%%%%%%%%%%%%%%%%%%

In this section we now investigate several statistical aspects dealing with the estimation of the number of species. We shall start with considering the problem of estimating the number of species, assuming first $\theta $ to be known. The proposed procedure to estimate $\left(\theta ,n\right)$ is explained after. Then we consider two stopping rules for the sampling process and a goodness-of-fit procedure. To conclude, numerical simulations were carried out. 

% % % % % % % % % % % % % % % % % % % % % % % % % % % % % % % % % % % % % % % 
\subsection{Estimation of $n$ when $\theta$ is known}
% % % % % % % % % % % % % % % % % % % % % % % % % % % % % % % % % % % % % % % 

Using theorems \ref{Theo6} and \ref{Theo6b}, one can easily derive the conditional distributions of $({\cal B}_{n,k}(1), \ldots, {\cal B}_{n,k}(p))$ and of $({\cal A}_{n,k}(1), \ldots, {\cal A}_{n,k}(k))$ which are respectively: 
\begin{equation}\label{e7}
{\bf P}_\theta({\cal B}_{n,k}(1) =b_1,\ldots,{\cal B}_{n,k}(p) =b_p\mid P_{n,k}=p) =\frac{k!}{p!}\frac{1}{B_{k,p}(\theta)}\prod_{q=1}^{p}\frac{(\theta)_{b_q}}{b_q!} 
\end{equation}
and:
\begin{equation}\label{e14}
{\bf P}_\theta ({\cal A}_{n,k}(1) =a_1,\ldots,{\cal A}_{n,k}(k)=a_k\mid P_{n,k}=p)=\frac{k!}{B_{k,p}(\theta)}\prod_{i=1}^{k}\frac{(\theta)_i^{a_i}}{i!^{a_i}a_i!} 
\end{equation}
These conditional probabilities being independent of $n$, it follows that the random variable $P_{n,k}$ is a sufficient statistic. 
\\[1ex]
Assume now that $k>p$. Using log-concavity in $n$ of ${\bf P}_\theta(P_{n,k}=p)$, the maximum likelihood estimator $\widehat{n}$ is given implicitly by:
\begin{equation*}
\frac{{\bf P}_\theta(P_{\widehat{n},k}=P)}{{\bf P}_\theta(P_{\widehat{n}-1,k}=P)}=1. 
\end{equation*}
From equation~(\ref{e6}), identifying $\widehat{n}$ with the largest integer short of the solution, the estimator $\widehat{n}$ we suggest is the fixed point of:
\begin{equation*}%\label{es1}
\widehat{n}=P+\widehat{n}\frac{((\widehat{n}-1)\theta) _{k}}{(\widehat{n}\theta ) _{k}} \;.  
\end{equation*}
This estimator is biased from above. The estimator $\widetilde{n}$ of Keener {\em et al.} (1986) is given by:
\begin{equation}\label{es2}
\widetilde{n}=P+\frac{B_{k,P-1}(\theta) }{B_{k,P}(\theta) } \;.  
\end{equation}

If $k\geqslant n$, it is unbiased attaining the minimum variance bound (UMVB) and in this case we have:
\begin{equation*}
{\bf E}_{\theta }\left( \frac{B_{k,P-1}(\theta)}{B_{k,P}(\theta)}\right) =n\frac{((n-1)\theta) _{k}}{(n\theta)_{k}}=n\left\langle \theta \right\rangle _{n,k;1}\;.
\end{equation*}
In practice, it is interesting to plot the observed number of species $P$ against sample size $k$. If $n<\infty $, $P$ should stabilize to an asymptote. If this is not the case, $P$ should drift at $\infty $ with $k$. For example, consider the following situation where $P/k\rightarrow \rho \in \left( 0,1\right) $ when $k\rightarrow \infty$ and $P\rightarrow \infty$. Using an asymptotic representation of $B_{k,p}(\theta)$ in this limit, one gets that $\tfrac{\widetilde{n}}{k} \rightarrow \rho _{*}$ where $\rho _{*}>0$ is defined implicitly (see Keener {\em et al.} 1986) by: 
\begin{equation*}
\rho =\rho _{*}\left(1-\left( \frac{\theta \rho _{*}}{1+\theta \rho _{*}} \right) ^{\theta }\right) \;.  
\end{equation*}
Asymptotic normality of $\left( \widetilde{n}-k\rho _{*}\right) /\sqrt{k}$ could be proved as $k\rightarrow \infty $.

\paragraph{Special cases}
\begin{enumerate}
\item {\bf Bose-Einstein case ($\theta=1$).} We find explicitly: 
\begin{equation*}
\widehat{n}=\frac{P(k-1)}{k-P} \qquad {\mbox{and}} \qquad 
\widetilde{n}=\frac{Pk}{k-P+1}. 
\end{equation*}
The maximal value which $\widehat{n}$ can take is obtained if $k-P=1$; in this case $\widehat{n}=P^{2}$. Its minimal value is $1$ if $P=1$ for all $k$. Note that in the Bose-Einstein model, $\rho _{*}=\rho /(1-\rho)$ and both $\widehat{n}/k$ and $\widetilde{n}/k$ would converge to $\rho_{*} $ assuming the asymptotic regime $P/k\rightarrow \rho $.

\item {\bf Maxwell-Boltzmann case ($\theta \rightarrow \infty $).} The maximum likelihood $\widehat{n}$ solves: 
\begin{equation*}
P=\widehat{n}\left(1-\left( 1-\frac{1}{\widehat{n}}\right) ^{k}\right) \;,
\end{equation*}
and, with $S_{k,p}$ the second kind Stirling numbers, the UMVB estimator $\widetilde{n}$ in (\ref{es2}) becomes: 
\begin{equation*}
\widetilde{n}=P+\frac{S_{k,P-1}}{S_{k,P}}\;, 
\end{equation*}
recalling $B_{k,p}(\theta) \sim \theta ^{k}S_{k,p}$ as $k\rightarrow \infty $.

\item {\bf Kingman case.} Indeed there is no estimation of $n$ problem (because $n=\infty $), rather the problem is to estimate $\gamma >0$ which is the unique remaining parameter. A situation in which the Kingman model fits best to data is a situation for which one should conclude $n=\infty $. Recalling equation~(\ref{e9}), the MLE $\widehat{\gamma }$ of $\gamma $ is characterized by $\partial _{\gamma }\log {\bf P}_{\gamma }^{*}(P_{k}=P)(\widehat{\gamma})=0$, hence implicitly by: 
\begin{equation}\label{es5}
\xi _{k}(\widehat{\gamma}) =\sum_{l=0}^{k-1}\frac{\widehat{\gamma}}{\widehat{\gamma}+l}=P \;.  
\end{equation}
It is biased and involves the problem of inverting the generalized harmonic sequence $\xi _{k}$. The properties of this estimator are well studied (see Carlton (1999) for a review). In particular, 
\begin{equation*}
\widehat{\gamma} \xrightarrow[k\rightarrow \infty]{Pr}\gamma \;.
\end{equation*}

In sharp contrast with the finite $n$ case, there is no UMVB estimator $\widetilde{\gamma }$ of $\gamma $ itself (nor of any polynomials in $\gamma $), because if $\widetilde{\gamma}=\phi(P)$ existed and were unbiased, function $\phi $ would satisfy:
\begin{equation*}
\sum_{p=1}^{k}\gamma^{p}s_{k,p}\phi (p) =\gamma(\gamma)_{k} \;, 
\end{equation*}
which is impossible because the left-hand-side is a polynomial of degree at most $k$ in $\gamma $ whereas the right-hand-side is a polynomial of degree $k+1$. So, if the problem is to estimate $\gamma$, $\widehat{\gamma}$ turns out to be the more satisfactory estimate in this case, despite its biased property. However there are UMVB estimators of rational functions of $\gamma$ of the form:
\begin{equation*}
\forall l \in \{1,\ldots,k \} \;, \quad r_{l}(\gamma) =\frac{\gamma(\gamma)_{k-l}}{(\gamma)_{k}} \;.  
\end{equation*}
They are given by:
\begin{equation}\label{es6}
\widetilde{r_{l}}=\frac{s_{k-l,P-1}}{s_{k,P}} \;. 
\end{equation}
For instance, from equation~(\ref{e9}), recalling $s_{k,p}=0$ if $p>k$:
\begin{equation*}
{\bf E}_{\gamma}^{*}\left(\frac{s_{k-l,P-1}}{s_{k,P}}\right) =\frac{\gamma}{(\gamma)_{k}}\sum_{p=1}^{k-l}\gamma^{p}s_{k-l,p}=r_{l}(\gamma) \;. 
\end{equation*}
In particular, when $l=1$, $\widetilde{r}_{1}=\tfrac{s_{k-1,P-1}}{s_{k,P}}$ is an UMVB estimator of $r_{1}(\gamma)=\tfrac{\gamma(\gamma)_{k-1}}{(\gamma)_{k}}=\tfrac{\gamma}{\gamma+k-1}$ which, from equation~(\ref{ex3}), is the probability to observe a new species from $k$-th trial (see Ewens, 1996).
\end{enumerate}

% % % % % % % % % % % % % % % % % % % % % % % % % % % % % % % % % % % % % % % 
\subsection{Joint estimations of $\theta$ and $n$} 
% % % % % % % % % % % % % % % % % % % % % % % % % % % % % % % % % % % % % % % 

In some applications, $\theta $ is also unknown and the question of its simultaneous estimation arises. As $P_{n,k}$ is not a sufficient statistic for $\theta $ (from equation~(\ref{e7}), for example), we turn to a different point of view. We briefly recall the idea of an estimator studied in Huillet and Paroissin (2005). With $(U_1,\ldots,U_k)$ the $k$ iid uniform random sample on $[0,1]$, let $(M_1,\ldots,M_k)$ be the corresponding fragments numbers hit (or animals species). With $l_1,l_2\in \{1,\ldots,k\}$, let:
\begin{equation*}
\delta _{l_{1},l_{2}}=\sum_{m=1}^{n}{\bf I}(M_{l_1}=m;M_{l_2}=m)  \;,
\end{equation*}
denote the event that $M_{l_1}=M_{l_2}$ for some fragment in ${\bf S}_{n}$. Introduce the pair-matching statistic: 
\begin{equation*}
D_{n,k}=\frac{1}{k(k-1) }\sum_{l_1\neq l_2=1}^{k}\delta _{l_1,l_2}\;.
\end{equation*}
It is the empirical probability that two randomly chosen items of the $k$-sample are identical. In a genetic context, $D_{n,k}$ is called the homozygosity of the sample (Tavar\'e, 2004). Note that:
\begin{equation*}
D_{n,k} = \frac{1}{k(k-1)}\sum_{q=1}^{P_{n,k}}{\cal B}_{n,k}(q)({\cal B}_{n,k}( q)-1) = \frac{k}{k-1}\left( \sum_{q=1}^{P_{n,k}}\left( \frac{{\cal B}_{n,k}\left(q\right) }{k}\right) ^{2}-\frac{1}{k}\right) \;.
\end{equation*}
Indeed for each visited species $q$, we need to count the number ${\cal B}_{n,k}(q)-1$ of returns to $q$, together with its multiplicity, with $\sum_{q=1}^{P_{n,k}}{\cal B}_{n,k}(q)=k$. Note that $D_{n,k}$ is a function of $P_{n,k}$ and of ${\cal B}_{n,k}(q)$, $q \in \{1,\ldots,P_{n,k}\}$.
\\[1ex]
The expectation of $\delta _{l_1,l_2}$ is the probability that two fragments chosen at random are the same. From equation~(\ref{e12}) with $k=2$ and $p=1$, $a_1=0$, $a_2=1$, we get:
\begin{equation}\label{es21}
{\bf E}_{\theta}(D_{n,k}) ={\bf E}_{\theta}(\delta_{l_1,l_2}) =\frac{1+\theta}{1+n\theta} \;.  
\end{equation}
Assume the observations are $P_{n,k}=P$ and ${\cal B}_{n,k}(q)=B_{q}$ for $q \in \{1,\ldots,P\}$. Then the observed value $D$ of $D_{n,k}$ is:
\begin{equation*}
D=\frac{k}{k-1}\left( \sum_{q=1}^{P}\left( \frac{B_{q}}{k}\right)
^{2}-\frac{1}{k}\right) \;. 
\end{equation*}
Applying the method of moments, $\theta$ can be estimated by $(1-D)/(nD-1)$ which is a consistent estimator. Therefore, we propose the following estimators of the pair $(\theta,n)$:
\begin{equation}\label{es22}
\widehat{\theta}_1=\frac{1-D}{\widehat{n}_1D-1} \quad {\mbox{and}} \quad \widehat{n}_1=P+\widehat{n}_1\frac{((\widehat{n}_1-1)\widehat{\theta}_1)_k}{(\widehat{n}_1\widehat{\theta}_1)_k} \;,  
\end{equation}
or:
\begin{equation}\label{es23}
\widetilde{\theta}_1=\frac{1-D}{\widetilde{n}_1D-1} \quad {\mbox{and}} \quad \widetilde{n}_1=P+\frac{B_{k,P-1}(\widetilde{\theta}_1)}{B_{k,P}(\widetilde{\theta_1})}\;. 
\end{equation}
These estimators are based on the couple of observations $(P,D)$ and depend on which estimator of $n$ itself was chosen. The numerical strategy is to get an implicit equation for $\widehat{n}_1$ (or $\widetilde{n}_1$) by substituting the expression of $\widehat{\theta}_1$ (or $\widetilde{\theta}_1$) as a function of $\widehat{n}_1$ (or $\widetilde{n}_1$) in the second equation, solve it in $\widehat{n}_1$ (or $\widetilde{n}_1$) as a fixed point problem and then deduce the corresponding estimates for $\theta$. Note that the functions involved in this fixed point problem are rational.
\\[2ex]
An alternative estimation procedure which uses the observable $P$ and cell occupancies ${\cal B}_{n,k}(q) =B_q$, $q \in \{1,\ldots,P\}$ is as follows. Consider the Renyi entropy of order $\alpha \geqslant 1$ (Pi\'elou, 1975) defined as follows:
\begin{equation*}
\phi_n = \frac{1}{1-\alpha} \log \sum_{m=1}^{n}S_m^\alpha \;. 
\end{equation*}
When $\alpha=2$, $\phi_n =-\log(\sum_{m=1}^{n}S_m^2)$ is the Simpson index of biodiversity (Simpson, 1949) up to the logarithmic transformation. As $\alpha$ tends to $1$, $\phi_n$ tends to $-\sum_{m=1}^{n}\log S_m$, the Shannon entropy, which is also an index of biodiversity (Pi\'elou, 1975). Consequently we will rather consider random additive functional of ${\bf S}_n$: 
\begin{equation*}
\phi_n =\sum_{m=1}^{n}h(S_m) \;. 
\end{equation*}
Hence with $h(s)=s^2$, it is exactly the Simpson index and with $h(s)=-s\log s$, it is the Shannon index. Below we will only consider the former case. The Simpson index of biodiversity can also be viewed as the size of a size-biased sample fragment from ${\bf S}_n$ for which:
\begin{equation*}
{\bf E}_{\theta}\left(\sum_{m=1}^{n}S_m^2\right) =n{\bf E}_{\theta}(S_1^2) = \frac{1+\theta}{1+n\theta} \;.
\end{equation*}
We obtain the same expected value as in equation~(\ref{es21}): it was already noticed by Simpson (1949). For a review of various measures of species diversity, see Hub\'alek (2000). According to lemma~1, ${\cal K}_{n,k}(m)/k$ is an estimator of $S_{m}$, implying  that the quantity:
\begin{equation*}
\widehat{\phi}_{n,k}=\sum_{m=1}^{n}h\left(\frac{{\cal K}_{n,k}(m)}{k}\right)
\end{equation*}
could be used as an estimator of $\phi_n$. Clearly $\widehat{\phi}_{n,k} \xrightarrow[k\rightarrow \infty]{d}\phi_n$ and in particular: 
\begin{equation*}
{\bf E}_{\theta}\left(\widehat{\phi}_{n,k}\right) \xrightarrow[k\rightarrow \infty]{} {\bf E}_{\theta}(\phi_n) \;. 
\end{equation*}
To be effective this supposes $n$ to be known, which could be not true. If $n$ is unknown, enhancing fragments with small probability to occur, we shall rather consider a quantity based on the sample coverage $C$ (the proportion of seen species in a $k$-sample):
\begin{equation*}
C=\sum_{q=1}^{P_{n,k}}S_{q} \;.
\end{equation*}
Indeed following Chao and Shen (2003), we can consider:
\begin{equation*}
\psi_{n,k} =\sum_{m=1}^{n}\frac{h\left( S_{m}^{\prime }\right) }{\pi _{m}\left(S_{m}^{\prime }\right) }{\bf I}\left( {\cal K}_{n,k}\left( m\right) >0\right) = \sum_{q=1}^{P_{n,k}}\frac{h\left( S_{q}^{\prime }\right) }{\pi _{q}\left(
S_{q}^{\prime }\right) } 
\end{equation*}
where $S_q^\prime= \tfrac{S_{q}}{C}$ for all $q \in \{1,\ldots,P_{n,k}\}$ (in order to have $\sum_{q=1}^{P_{n,k}}S_q^\prime=1$) and where $\pi _{q}( S_{q}^{\prime })=1-( 1-S_{q}^{\prime}) ^{k}$ is the probability to observe fragment $q$ among the $P_{n,k}$ which were effectively observed (see equation~(\ref{e4})). Clearly, we have:
\begin{equation*}
\psi_{n,k} \xrightarrow[k\rightarrow \infty]{d} \phi_n \;.
\end{equation*}
But $\psi_{n,k}$ involves unknown quantities. Hence, if $P_{n,k}$ fragments are observed, an estimator of $S_q^\prime=S_{q}/C$ is: 
\begin{equation*}
\widehat{S}_q^\prime=\frac{{\cal B}_{n,k}(q)}{k} \;. 
\end{equation*}
It follows that a possible estimator of $\psi_{n,k}$ is: 
\begin{equation*}
\widehat{\psi}_{n,k}=\sum_{q=1}^{P_{n,k}}\frac{h\left( \frac{{\cal B}_{n,k}(q)}{k}\right) }{1-\left( 1-\frac{{\cal B}_{n,k}(q)}{k}\right) ^{k}} \;. 
\end{equation*}
Particularizing to $h(s) =s^2$ (Simpson index of diversity), 
\begin{equation*}
\widehat{\psi}_{n,k}=\sum_{q=1}^{P_{n,k}}\frac{\left(\frac{{\cal B}_{n,k}(q)}{k}\right)^2}{1-\left(1-\frac{{\cal B}_{n,k}(q)}{k}\right)^{k}} \;, 
\end{equation*}
is such that: 
\begin{equation*}
{\bf E}_{\theta}(\widehat{\psi}_{n,k})\xrightarrow[k\rightarrow \infty]{} {\bf E}_{\theta}(\phi_n) = \frac{\theta +1}{n\theta +1} \;. 
\end{equation*}
Assuming the observation is $\widehat{\psi}_{n,k}=\psi$, $(1-\psi)/(n\psi-1)$ is also a consistent estimator of $\theta$ by application of the asymptotic method of moments. Note that: 
\begin{equation*}
\psi =\sum_{q=1}^{P}\frac{\left( \frac{B_q}{k}\right)^2}{1-\left(1-\frac{B_q}{k}\right)^k} \;, 
\end{equation*}
where $B_q \geqslant 1$ is an observed realization of ${\cal B}_{n,k}(q)$ for any $q \in \{1,\ldots,P\}$. hence it involves the observations $P$ and $B_{q}$ with $q \in \{1,\ldots,P \}$. Therefore an alternative closely related to the two previous estimators (see equations~(\ref{es22}) and (\ref{es23})) for the pair $(\theta,n)$ could be: 
\begin{equation*}
\widehat{\theta}_2=\frac{1-\psi}{\widehat{n}_2\psi-1} \quad {\mbox{and}} \quad \widehat{n}_2=P+\widehat{n}_2\frac{((\widehat{n}_2-1)\widehat{\theta}_2)_k}{(\widehat{n}_2\widehat{\theta}_2)_k} \;,
\end{equation*}
or:
\begin{equation*}
\widetilde{\theta}_2=\frac{1-\psi}{\widetilde{n}_2\psi-1} \quad {\mbox{and}} \quad \widetilde{n}_2=P+\frac{B_{k,P-1}(\widetilde{\theta}_2)}{B_{k,P}(\widetilde{\theta}_2)} \;.  
\end{equation*}
These estimators are based on the set of observations $P$ and $B_q$, $q \in \{1,\ldots,P \}$ and depend on which estimator of $n$ itself was chosen.

% % % % % % % % % % % % % % % % % % % % % % % % % % % % % % % % % % % % % % %
\subsection{Stopping rules}
% % % % % % % % % % % % % % % % % % % % % % % % % % % % % % % % % % % % % % %

Here we now define three stopping rules for the sampling process. Indeed there is no "objective" stopping rules. Each of these stopping rules are based on some simple and interesting questions that arise naturally in our context. What is the sample size until the first visit to the smallest fragment? How long should one wait until all fragments have been visited (the coupon collector problem)? When the probability to discover a new species is smaller that a given threshold? Clearly the two first questions concern only the situation with $n<\infty$ while the last could be answered for any case.

\begin{enumerate}

\item Let $S_{\left( n\right) }$ be the smallest fragment among ${\bf S}_{n}$. Let $K_{\left( n\right) }$ be the sample size until the first visit to $S_{\left( n\right) }$. Then 
\begin{equation*}
{\bf P}\left( K_{\left( n\right) }>k\mid {\bf S}_{n}\right) =\left(
1-S_{\left( n\right) }\right) ^{k}  
\end{equation*}
is the conditional waiting time until the first visit to this fragment. Averaging over the partitions ${\bf S}_{n}$, we obtain 
\begin{eqnarray*}
{\bf P}_{\theta }\left( K_{\left( n\right) }>k\right) 
 &= & {\bf E}_{\theta }{\bf P}\left( K_{\left( n\right) }>k\mid {\bf S}_{n}\right) ={\bf E}_{\theta}\left[ \left( 1-S_{\left( n\right) }\right) ^{k}\right] \\
 & = &\int_{0}^{1}{\bf P}_{\theta }\left[ \left( 1-S_{\left( n\right) }\right)^{k}>s\right] ds=1-\int_{0}^{1}{\bf P}_{\theta }\left[ S_{\left( n\right)}>1-s^{\frac{1}{k}}\right] ds.
\end{eqnarray*}
To evaluate this probability, we thus need to compute the distribution of $S_{\left(n\right)}$. We can prove 
\begin{equation*}
{\bf P}_{\theta }\left( S_{\left( n\right) }>s\right) =\frac{\Gamma \left(
n\theta \right) }{\Gamma \left( \theta \right) ^{n}}\phi _{n,\theta }\left(
s\right)  
\end{equation*}
where $\phi_{n,\theta}(s)=h_{\theta,s}^{*n}(t)\mid_{t=1}$ is the $n$-fold convolution of $t \mapsto h_{\theta,s}(t)=t^{\theta-1}{\bf I}(1\geqslant t>s)$ evaluated at $t=1$. This distribution could be computed in closed form. In the Bose-Einstein case ($\theta =1$), the expression simplifies to 
\begin{equation*}
{\bf P}_{1}(S_{(n)}>s)=\left(1-ns\right)_{+}^{n-1} \;,
\end{equation*}
where $x_{+}=x\vee 0$ (see Huillet, 2003). As a result, with $k\geqslant 1$, 
\begin{eqnarray*}
{\bf P}_{1}\left( K_{n}>k\right) 
 & = & 1-\int_{0}^{1}\left( 1-n\left(1-s^{1/k}\right) \right) _{+}^{n-1}ds \\ 
 & = & 1-\int_{\left( 1-\frac{1}{n}\right) ^{k}}^{1}\left( 1-n\left(1-s^{1/k}\right) \right) ^{n-1}ds \\ 
 & = & 1-\frac{k}{n}\left( 1-\frac{1}{n}\right) ^{k-1}\int_{0}^{1}x^{n-1}\left( 1+\frac{x}{n-1}\right) ^{k-1}dx \\ 
 & = & 1-\frac{k}{n}\left( 1-\frac{1}{n}\right) ^{k-1}\sum_{j=0}^{k-1}\binom{k-1}{j}\left( n-1\right) ^{-j}\left( n+j\right) ^{-1} \;.
\end{eqnarray*}

\item Let $K_n^+=\inf\{ k\geqslant n \;;\; P_{n,k}=n\}$ be the first time that all species are observed in the sample. With $k\geqslant n$, we have 
\begin{equation*}
{\bf P}_{\theta}(K_n^+>k)={\bf P}_{\theta}(P_{n,k}<n)=1-{\bf P}_{\theta}(P_{n,k}=n) \;. 
\end{equation*}
Recalling equation~(\ref{as3}), we obtain 
\begin{equation*}
{\bf P}_{\theta}(K_n^+>k)=\frac{1}{(n\theta)_k}\sum_{q=1}^{n-1}(-1)^{q-1}\binom{n}{q}((n-q)\theta)_k \;.  
\end{equation*}
Recalling equation~(\ref{e6}), with $k\geqslant n$, this may also be written as follows:
\begin{equation*}
{\bf P}_{\theta}(K_n^+\leqslant k)=\frac{n!}{(n\theta)_k}B_{k,n}(\theta)  \;. 
\end{equation*}
When $\theta=1$ and $\theta\uparrow\infty$, this formula further simplifies to give respectively the well-known results:  
\begin{equation*}
{\bf P}_{1}(K_n^+\leqslant k) = \frac{\binom{k}{n}}{\binom{n+k-1}{n}}=\frac{\binom{k-1}{n-1}}{\binom{n+k-1}{k}} \;, 
\end{equation*}
and:
\begin{equation*}
{\bf P}_{\infty}(K_n^+\leqslant k)=\frac{n!S_{k,n}}{n^{k}} \;.
\end{equation*}

\item A last possible stopping rule for the sample is the following: the proceeding with sampling is useless if the estimated probability to obtain a new species $\widehat{r}_1$ (or $\widetilde{r}_1$) is less than some small value $\varepsilon$ (say $\varepsilon=0.01$ for instance). Hence we are interested in the two following sample sizes:
\begin{equation*}
\widehat{K}_\varepsilon=\inf \{ k \;;\; \widehat{r}_1<\varepsilon\} 
\quad {\mbox{and}} \quad  
\widetilde{K}_\varepsilon=\inf \{ k \;;\; \widetilde{r}_1<\varepsilon\}  \;,
\end{equation*}
using respectively $\widehat{n}$ or $\widetilde{n}$ for estimating $n$. When $P$ distinct species have been observed at step $k$, the probability to get a new species at the $(k+1)$-th trial is 
\begin{equation*}
r_1=\frac{(n-P)\theta}{n\theta+k} \;.  
\end{equation*}
Using estimators developed previously, we obtain the two following estimates for $r_1$:
\begin{equation*}
\widehat{r}_1=\frac{(\widehat{n}-P)\theta}{\widehat{n}\theta+k} 
\quad {\mbox{and}} \quad  
\widetilde{r}_1=\frac{(\widetilde{n}-P)\theta}{\widetilde{n}\theta+k} 
\end{equation*}
(if $\theta$ is also unknown, one could replace its estimates). When $\theta=1$ (Bose-Einstein case), the explicit expressions for $\widehat{n}$ and for $\widetilde{n}$ lead to:
\begin{equation*}
\widehat{r}_1=\frac{\widehat{n}-P}{\widehat{n}+k}=\frac{P(P-1)}{k^2-P}  
\quad {\mbox{and}} \quad  
\widetilde{r}_1=\frac{\widehat{n}-P}{\widehat{n}+k}=\frac{P(P-1)}{k^2+1} \;.
\end{equation*}
Obviously $\widehat{r}_1 \geqslant \widetilde{r}_1$. As a consequence, $\widehat{K}_{\varepsilon} \geqslant \widetilde{K}_{\varepsilon}$. Thus, $\widehat{K}_{\varepsilon}$ is of order $P\varepsilon^{-1/2}$ and $\widetilde{K}_{\varepsilon}$ is of order $(P(P-1))^{1/2}\varepsilon^{-1/2}$. Hence if $P$ is large enough, $\widehat{K}_{\varepsilon}$ and $\widetilde{K}_{\varepsilon}$ are of the same order. In the case of Kingman model, we have only an explicit expression for $\widetilde{n}$. In such case, let us recall that:
\begin{equation*}
\widetilde{r}_{1}=\frac{s_{k-1,P-1}}{s_{k,P}} \;,
\end{equation*}
which could be evaluated from inspection of a table of the first kind Stirling numbers.
\end{enumerate}

% % % % % % % % % % % % % % % % % % % % % % % % % % % % % % % % % % % % % % %
\subsection{Goodness of fit using the second Ewens sampling formula}
% % % % % % % % % % % % % % % % % % % % % % % % % % % % % % % % % % % % % % %

Deciding which model fits the best to a concrete situation is a challenging problem. This can first be appreciated from the likelihood of the observations under the different models to be compared. We shall recall an additional procedure followed by Keener {\em et al.} (1987) for the case $n<\infty $: First, a simple computation of $\alpha _{i}={\bf E}_\theta({\cal A}_{n,k}(i))$ gives, using our notations 
\begin{equation*}
\alpha _{i}=ni\binom{k}{i}\left\langle \theta \right\rangle _{n,k-i;1} \;.
\end{equation*}
According to theorem~2.5 in Keener {\em et al.} (1987), a UMVB estimator of $\alpha_i$ is obtained under the form: 
\begin{equation*}
\widetilde{\alpha}_i={\bf E}_\theta({\cal A}_{n,k}(i) \mid P) =\frac{(\theta)_i}{(i-1)!}\frac{B_{k-i,P-1}(\theta)}{B_{k,P}(\theta)} \;.
\end{equation*}
When $\theta=1$ (Bose-Einstein case), it becomes: 
\begin{equation*}
\widetilde{\alpha}_i=\frac{ip(p-1)}{k-i+1}\frac{(k-i-1)!(k-p)!}{k!(k-1)!} \;, 
\end{equation*}
recalling the expression of $B_{k,P}(1)$. Define next the MLE $\widehat{\alpha}_i$ of $\alpha_i$ to be: 
\begin{equation*}
\widehat{\alpha }_{i}=\widehat{n}i\binom{k}{i}\left\langle \theta
\right\rangle _{\widehat{n},k-i;1}. 
\end{equation*}
Based on the observations $A_{i}$ of ${\cal A}_{n,k}(i)$, the goodness of fit of the model can be measured by one of the two following statistics: 
\begin{equation*}
\widetilde{\chi}^2 = \sum_{i=1}^k (A_i-\widetilde{\alpha}_{i})^2/\widetilde{\alpha}_i \qquad {\mbox{or}} 
\qquad 
\widehat{\chi}^2=\sum_{i=1}^k (A_i-\widehat{\alpha}_i)^2/\widehat{\alpha}_i \;.
\end{equation*}
In the case of the Kingman model, one can check that:
\begin{equation*}
\alpha_i={\bf E}_\gamma^*({\cal A}_k(i))=\frac{k!}{i(k-i)!}\frac{\gamma}{\gamma+i-1}  
\end{equation*}
and the second statistic becomes: 
\begin{equation*}
\widehat{\chi}^2=\sum_{i=1}^k(A_i-\widehat{\alpha}_i)^2/\widehat{\alpha}_i \;,
\end{equation*}
where $\widehat{\alpha}_i=\tfrac{k!}{(k-i)!}\frac{\widehat{\gamma}}{i(\widehat{\gamma}+i-1)}$ and where $\widehat{\gamma}$ is given by equation~(\ref{es5}).

% % % % % % % % % % % % % % % % % % % % % % % % % % % % % % % % % % % % % % % 
\subsection{Numerical simulations}
% % % % % % % % % % % % % % % % % % % % % % % % % % % % % % % % % % % % % % % 

We now apply the estimators developed and discussed previously on simulated data to observe the behavior of their quality when $n$, $\theta$ and $k$ are varying. 
\\[1ex]
We consider empirical distribution of $k$-samples for a given Dirichlet partition (however one could rather prefer to consider empirical distribution of Dirichlet partitions and one $k$-sample). For $n\in \{100;200;500\}$ and $\theta \in \{\tfrac{1}{2};1;\tfrac{3}{2}\}$, we simulated a Dirichlet partition. Over this partition, we simulated 500 $k$-samples with $k\in \{\tfrac{2n}{3};n;\tfrac{3n}{2}\}$. Note that we managed to use the same uniform random variables, so that $P_{n,\tfrac{3n}{2}}$ corresponds to the same observations than the first ones of $P_{n,n}$ and so on. We considered the two different cases: $\theta$ known and $\theta$ unknown. In some cases, we did not use the estimators based on $\widetilde{n}$ since computations were too heavy. 
\\[1ex]
Tables \ref{tab:simu1} to \ref{tab:simu3} contain the estimations (first when $\theta$ is known and then when $\theta$ is unknown) respectively for $\theta =1$, $\theta =\frac{1}{2}$ and $\theta =\tfrac{3}{2}$. The numbers that appear in the cells are the empirical averages of the estimations of $n$ or $\theta $, and the number in brackets within the cells the empirical standard deviations (over 100 $k$-samples as described above). Note that comparing standard deviations for the estimations of $n$ and $\theta $ does not make sense (one should rather use for instance the coefficient of variation which is dimension-less). 
\\[1ex]
For $\theta =1$ (table \ref{tab:simu1}), the results are very good, even when $\theta $ is unknown (but except for the estimation of $\theta $ with the statistic $\Psi _{n,k}$). For $\theta =\frac{1}{2}$ and $\theta =\tfrac{3}{2}$, we did not run the estimators based on $\widetilde{n}$ for the reason given above. Results are quite good when $\theta $ is known, but not so good when $\theta $ is unknown.

\begin{table}[htbp]
\begin{center}
\begin{tabular}{|cc|cc|cccc|}
\hline
&  & \multicolumn{2}{c}{$\theta$ known} & \multicolumn{4}{|c|}{$\theta$ unknown} \\ 
&  & $\widehat{n}$ & $\widetilde{n}$ & \multicolumn{2}{c}{$(\widehat{n}_1,\widehat{\theta}_1)$ based on $D_{n,k}$} & \multicolumn{2}{c|}{$(\widehat{n}_2,\widehat{\theta}_2)$ based on $\psi_{n,k}$} \\ 
\hline\hline
$n=100$ & $k=66$ & {\bf 92.96} & 90.98 & 1.2E+09 & 0.21 & 9.8E+08 & 2.7E-08 \\ 
&  & (16.51) & (15.80) & (1.3E+09) & (3.5E+08) & (0.32) & (1.1E-08) \\ 
& $k=100$ & {\bf 93.69} & {\bf 92.81} & 2821.86 & 0.21 & 3322.34 & 0.01 \\ 
&  & (13.29) & (13.05) & (1884.31) & (12632.13) & (0.33) & (0.01) \\ 
& $k=150$ & {\bf 91.83} & 91.46 & {\bf 91.83} & {\bf 1.13} & 130.06 & {\bf 0.61} \\ 
&  & (9.62) & (9.55) & (9.63) & (268.89) & (0.23) & (0.12) \\ 
\hline\hline
$n=200$ & $k=133$ & 206.76 & {\bf 204.33} & {\bf 223.08} & {\bf 1.09} &  4731.03 & 0.01 \\ 
&  & (31.23) & (30.46) & (153.28) & (0.23) & (767.15) & (0.001) \\ 
& $k=200$ & {\bf 201.22} & {\bf 200.21} & {\bf 201.22} & {\bf 1.11} & {\bf 201.22} & 0.46 \\ 
&  & (18.07) & (17.89) & (18.07) & (0.18) & (18.07) & (0.04) \\ 
& $k=300$ & {\bf 200.30} & {\bf 199.85} & {\bf 200.30} & {\bf 1.09} & {\bf 200.30} & 0.58 \\ 
&  & (13.28) & (13.22) & (13.28) & (0.15) & (13.28) & (0.05) \\ 
\hline\hline 
$n=500$ & $k=333$ & {\bf 513.34} & {\bf 510.96} & {\bf 513.34} & {\bf 0.99}
& {\bf 513.34} & 0.32 \\ 
&  & (41.89) & (41.49) & (41.89) & (0.13) & (41.89) & (0.02) \\ 
& $k=250$ & {\bf 515.20} & {\bf 514.14} & {\bf 515.20} & {\bf 0.99} & {\bf 515.20} & 0.43 \\ 
&  & (27.79) & (27.68) & (27.79) & (0.13) & (27.79) & (0.03) \\ 
& $k=750$ & {\bf 515.15} & {\bf 514.68} & {\bf 515.15} & {\bf 0.98} & {\bf 515.15} & 0.54 \\ 
&  & (21.61) & (21.57) & (21.61) & (0.10) & (21.61) & (0.03) \\ 
\hline
\end{tabular}
\end{center}
\caption{Estimation over simulated data with $\theta=1$}
\label{tab:simu1}
\end{table}

\begin{table}[htbp]
\begin{center}
\begin{tabular}{|cc|c|cccc|}
\hline
&  & $\theta$ known & \multicolumn{4}{|c|}{$\theta$ unknown} \\ 
&  & $\widehat{n}$ & \multicolumn{2}{c}{$(\widehat{n}_1,\widehat{\theta}_1)$ based on $D_{n,k}$} & \multicolumn{2}{c|}{$(\widehat{n}_2,\widehat{\theta}_2)$ based on $\psi_{n,k} $} \\ 
\hline\hline
$n=100$ & $k=66$ & {\bf 108.21} & 8.2E+08 & 0.16 & 7.5E+08 & 3.2E-08 \\ 
&  & (19.39) & (8.5E+08) & (0.30) & (2.7E+08) & (1.3E-08) \\ 
& $k=100$ & {\bf 110.36} & 1418.81 & {\bf 0.29} & 1708.36 & 0.02 \\ 
&  & (13.93) & (1547.26) & (0.32) & (968.16) & (0.02) \\ 
& $k=150$ & 162.34 & {\bf 156.51} & 0.91 & 2462.54 & {\bf 0.38} \\ 
&  & (15.58) & (614.64) & (0.23) & (16084.16) & (0.29) \\ 
\hline\hline
$n=200$ & $k=133$ & 136.75 & 4.2E+08 & 0.33 & 5.9E+08 & 0.002 \\ 
&  & (84.83) & (6.9E+08) & (0.41) & (3.8E+08) & (0.005) \\ 
& $k=200$ & 141.06 & 836.19 & {\bf 0.42} & 1222.67 & 0.11 \\ 
&  & (87.25) & (1328.58) & (0.38) & (1371.98) & (0.19) \\ 
& $k=300$ & {\bf 183.38} & {\bf 473.93} & {\bf 0.69} & 2828.47 & 0.19 \\ 
&  & (69.20) & (3672.28) & (0.310) & (30533.87) & (0.26) \\ 
\hline\hline
$n=500$ & $k=333$ & 232.11 & 4.2E+08 & 0.35 & 5.9E+08 & 0.07 \\ 
&  & (270.59) & (6.9E+08) & (0.39) & (3.8E+08) & (0.15) \\ 
& $k=250$ & 234.31 & 882.81 & {\bf 0.41} & 1269.30 & 0.11 \\ 
&  & (272.37) & (1308.03) & (0.37) & (1338.81) & (0.18) \\ 
& $k=750$ & {\bf 277.95} & {\bf 521.22} & {\bf 0.68} & 2875.75 & 0.19 \\ 
&  & (256.56) & (3669.47) & (0.30) & (30529.92) & (0.25) \\ 
\hline
\end{tabular}
\end{center}
\caption{Estimation over simulated data with $\theta=\tfrac{1}{2}$}
\label{tab:simu2}
\end{table}

\begin{table}[htbp]
\begin{center}
\begin{tabular}{|cc|c|cccc|}
\hline
&  & $\theta$ known & \multicolumn{4}{|c|}{$\theta$ unknown} \\ 
&  & $\widehat{n}$ & \multicolumn{2}{c}{$(\widehat{n}_1,\widehat{\theta}_1)$ based on $D_{n,k}$} & \multicolumn{2}{c|}{$(\widehat{n}_2,\widehat{\theta}_2)$ based on $\psi_{n,k} $} \\ 
\hline\hline
$n=100$ & $k=66$ & {\bf 99.39} & 7.801E+08 & 0.18 & 8.299E+08 & 0.0001 \\ 
&  & (19.49) & (1.167E+09) & (0.28) & (3.365E+08) & (0.003) \\ 
& $k=100$ & {\bf 99.37} & 1740.15 & 0.26 & 1674.75 & 0.01 \\ 
&  & (13.48) & (2273.91) & (0.30) & (1331.18) & (0.01) \\ 
& $k=150$ & 155.01 & {\bf 465.64} & {\bf 0.74} & 2820.17 & 0.0001 \\ 
&  & (18.25) & (3673.02) & (0.37) & (30534.60) & (0.25) \\ 
\hline\hline
$n=200$ & $k=133$ & 151.57 & 4.195E+08 & 0.40 & 5.922E+08 & 0.002 \\ 
&  & (110.93) & (6.926E+08) & (0.47) & (3.830E+08) & (0.005) \\ 
& $k=200$ & 151.08 & 853.30 & 0.47 & 1239.78 & 0.09 \\ 
&  & (107.15) & (1320.33) & (0.45) & (1359.18) & (0.15) \\ 
& $k=300$ & {\bf 189.47} & {\bf 488.62} & {\bf 0.75} & 2843.15 & 0.19 \\ 
&  & (81.03) & (3671.15) & (0.38) & (30532.61) & (0.25) \\ 
\hline
$n=500$ & $k=333$ & 254.38 & 4.2E+08 & 0.37 & 5.9E+08 & 0.05 \\ 
&  & (314.97) & (6.9E+08) & (0.41) & (3.8E+08) & (0.11) \\ 
& $k=250$ & 253.11 & 921.32 & 0.44 & 1307.81 & 0.09 \\ 
&  & (309.82) & (1295.86) & (0.40) & (1315.75) & (0.15) \\ 
& $k=750$ & {\bf 291.79} & {\bf 556.83} & {\bf 0.71} & 2911.36 & 0.18 \\ 
&  & (284.10) & (3668.96) & (0.33) & (30527.13) & (0.24) \\ 
\hline
\end{tabular}
\end{center}
\caption{Estimation over simulated data with $\theta=\frac{3}{2}$}
\label{tab:simu3}
\end{table}

%%%%%%%%%%%%%%%%%%%%%%%%%%%%%%%%%%%%%%%%%%%%%%%%%%%%%%%%%%%%%%%%%%%%%%%%%%%%%%%
\section{Applications to real data}
%%%%%%%%%%%%%%%%%%%%%%%%%%%%%%%%%%%%%%%%%%%%%%%%%%%%%%%%%%%%%%%%%%%%%%%%%%%%%%%

We applied our estimators to fourteen different sets of real data. These data are of various nature as we will see later. However we will only consider here two kind of real data sets. The first one was studied by Keener {\em eta al.} (1987) and deals with word usage by two different authors. The interest of this first data set is that we indeed known the number $n$ to be estimated. The second one was extracted from observations made by Janzen (1973): these data correspond to beetles species observed either during the day or during the night and at different season.

% % % % % % % % % % % % % % % % % % % % % % % % % % % % % % % % % % % % % % %
\subsection{Federalist papers data}
% % % % % % % % % % % % % % % % % % % % % % % % % % % % % % % % % % % % % % %

These data were considered by Mosteller and Wallace (1984) and concern word usage by James Madison and Alexander Hamilton. {\em The Federalist} papers were written between 1787 and 1788 to promote the new Constitution of the State of New-York. Published in various newspapers, these papers was signed under a pseudonym (as for instance {\em Publius}). Each paper was written by one of the three following persons: James Madison, Alexander Hamilton and John Jay.  The author of most of the seventy-seven papers is clearly identified but Madison and Hamilton disputed the authorship of twelve of them. Hence in order to determine the author of these disputed papers, many researchers studied papers written surely by Madison and Hamilton. In particular some of them focused on the occurrences of function words as defined by the Miller-Newman-Friedman list. Mosteller and Wallace (1984) developed a Bayesian approach to solve this problem. In order to do so, they divided a set of well identified texts (either by Madison or by Hamilton) of equal length. It corresponds to the data presented below.
\\[1ex]
For $q \in \{0, \ldots, 6\}$, the two tables below gives the number $A(q)$ of manuscripts (of the same type as those published in {\em The Federalist} and with comparable length) in which a specific word ('may' for Madison and 'can' for Hamilton) occurs exactly $q$ times (these two words were the one selected by Keener {\em et al.} (1987), but Mosteller and Wallace (1984) studied more words). In this case, the exact numbers of manuscripts are known (respectively $n=262$ and $n=247$) and so we will be able to compare it with our estimation. Since we have this additional information, tables~\ref{tab:real-1-1} and~\ref{tab:real-1-2} contain the column for $q=0$, which is not available in real applied context.
\\[1ex]
For these two data sets, Keener {\em et al.} (1987) computed the estimation of $n$ for the three special cases considered here (i.e. for the Bose-Einstein, Maxwell-Boltzmann and the Kingman models) and also for the case where both $n$ and $\theta$ are unknown. Results are compared throughout the log-likelihood. Indeed, in their paper, there is no theoretical development when $n$ and $\theta$ are unknown.

\begin{itemize}

\item {\em Madison data:}  the sample size is $k=172$ and the number of distinct kinds of manuscripts is $p=106$. When using the statistic $D_{n,k}$, we obtain $\widehat{n}_1=274.6$ and $\widehat{\theta}_1=1.09$. When using the statistic $\widehat{\psi}_{n,k}$, we obtain $\widehat{n}_2=274.6$ and $\widehat{\theta}_2=0.32$. When assuming that both $n$ and $\theta$ are unknown, Keener {\em et al.} (1987) obtained respectively 217 and 1.998 as estimated values. This value for $n$ is far from its correct value.

\begin{table}[ht]
\begin{center}
\begin{tabular}{|c||c|c|c|c|c|c|c|}
\hline
$q$ & 0 & 1 & 2 & 3 & 4 & 5 & 6  \\ 
$A(q)$ & 156 & 63 & 29 & 8 & 4 & 1 & 1  \\ 
\hline
\end{tabular}
\end{center}
\caption{Madison data}
\label{tab:real-1-1}
\end{table}

\item {\em Hamilton data:} the sample size is $k=139$ and the number of distinct kinds of manuscripts is $p=90$. When using the statistic $D_{n,k}$, we obtain $\widehat{n}_1=253.5$ and $\widehat{\theta}_1= 0.85$. When using the statistic $\widehat{\psi}_{n,k}$, we obtain $\widehat{n}_2=4526.3$ and $\widehat{\theta}_2=0.01$. The second value is unsatisfactory. However when assuming that both $n$ and $\theta$ are unknown, Keener {\em et al.} (1987) obtained respectively 10,000,001 and $1.094 \times 10^{-5}$ as estimated values! This value for $n$ is strongly far from its correct value. 

\begin{table}[ht]
\begin{center}
\begin{tabular}{|c||c|c|c|c|c|c|c|}
\hline
$q$ & 0 & 1 & 2 & 3 & 4 & 5 & 6  \\ 
$A(q)$ & 157 & 60 & 20 & 5 & 2 & 2 & 1  \\ 
\hline
\end{tabular}
\end{center}
\caption{Hamilton data}
\label{tab:real-1-2}
\end{table}
\end{itemize}

% % % % % % % % % % % % % % % % % % % % % % % % % % % % % % % % % % % % % % %
\subsection{Tropical insect data}
% % % % % % % % % % % % % % % % % % % % % % % % % % % % % % % % % % % % % % %

Janzen (1973) observed tropical insects in twenty-five different sites in Costa Rica and the Caribbean Islands. This paper contains a remarkable collection of data. From it, we extracted three series corresponding to beetles collected either in day-time or in night-time, all during a dry season. These data were collected at the same site referred as "Osa secondary" in Janzen (1973). Observations of the first series were collected during the dry season of the year 1967 in day-time while the ones of the second series were collected at the same period in night-time. At least observations of the last series were collected during the dry season of the year 1968 in day-time.

\begin{itemize}

\item {\em Osa secondary/day/dry/1967: } it was observed $k=996$ beetles and $p=140$ distinct species. When using the statistic $D_{n,k}$, we obtain $\widehat{n}_1=162.7$ and $\widehat{\theta}_1=0.219$. When using the statistic $\widehat{\psi}_{n,k}$, we obtain $\widehat{n}_2=162.7$ and $\widehat{\theta}_2=0.211$. 

\begin{table}[ht]
\begin{center}
\begin{tabular}{|c||c|c|c|c|c|c|c|c|c|c|c|c|c|}
\hline
$q$ & 1 & 2 & 3 & 4 & 5 & 6 & 7 & 8 & 9 & 10 & 11 & 12 & 14 \\  
$A(q)$ & 70 & 17 & 4 & 5 & 5 & 5 & 5 & 3 & 1 & 2 & 3 & 2 & 2  \\   
\hline
\hline
$q$ & 17 & 29 & 20 & 21 & 24 & 26 & 40 & 57 & 60 & 64 & 71 & 77 & \\  
$A(q)$ & 1 & 2 & 3 & 1 & 1 & 1 & 1 & 2 & 1 & 1 & 1 & 1 & \\   
\hline
\end{tabular}
\end{center}
\caption{Osa secondary/day/dry/1967 data}
\label{tab:janzen-1}
\end{table}

\item {\em Osa secondary/night/dry/1967: } it was observed $k=835$ beetles and $p=151$ distinct species. When using the statistic $D_{n,k}$, we obtain $\widehat{n}_1=184.1$ and $\widehat{\theta}_1=0.268$. When using the statistic $\widehat{\psi}_{n,k}$, we obtain $\widehat{n}_2=184.1$ and $\widehat{\theta}_2=0.252$. 

\begin{table}[ht]
\begin{center}
\begin{tabular}{|c||c|c|c|c|c|c|c|c|c|c|c|}
\hline
$q$ & 1 & 2 & 3 & 4 & 5 & 7 & 8 & 9 & 10 & 11 & 12 \\
$A(q)$ & 61 & 24 & 13 & 12 & 5 & 6 & 5 & 2 & 4 & 2 & 3 \\
\hline
\hline
$q$ & 13 & 15 & 17 & 18 & 19 & 26 & 30 & 33 & 40 & 44 & 62 \\
$A(q)$ & 1 & 1 & 1 & 2 & 2 & 1 & 1 & 1 & 1 & 1 & 2 \\
\hline
\end{tabular}
\end{center}
\caption{Osa secondary/night/dry/1967 data}
\label{tab:janzen-2}
\end{table}

\item {\em Osa secondary/day/dry/1968: } it was observed $k=807$ beetles and $p=143$ distinct species. When using the statistic $D_{n,k}$, we obtain $\widehat{n}_1=173.6$ and $\widehat{\theta}_1=0.111$. When using the statistic $\widehat{\psi}_{n,k}$, we obtain $\widehat{n}_2=173.6$ and $\widehat{\theta}_2=0.108$. 

\begin{table}[ht]
\begin{center}
\begin{tabular}{|c||c|c|c|c|c|c|c|c|c|c|c|c|}
\hline
$q$ & 1 & 2 & 3 & 4 & 5 & 6 & 7 & 9 & 10 & 11 & 12 & 13 \\
$A(q)$ & 85 & 12 & 10 & 4 & 6 & 3 & 5 & 1 & 2 & 1 & 1 & 1 \\
\hline
\hline
$q$ & 15 & 18 & 20 & 24 & 25 & 28 & 29 & 30 & 79 & 106 & 112 & \\
$A(q)$ & 1 & 2 & 1 & 1 & 1 & 1 & 1 & 1 & 1 & 1 & 1 & \\
\hline
\end{tabular}
\end{center}
\caption{Osa secondary/day/dry/1968}
\label{tab:janzen-3}
\end{table}

\end{itemize}

For all the three data sets, the two values of $\widehat{n}$ are identical. Moreover the two values of $\widehat{\theta}$ are close, which is not always the case. It may be due to the fact that many species are abundant.

% % % % % % % % % % % % % % % % % % % % % % % % % % % % % % % % % % % % % % %
\subsection{Conclusion}
% % % % % % % % % % % % % % % % % % % % % % % % % % % % % % % % % % % % % % %

These two families of data sets give some illustration of the results obtained when applying the estimators developed in this paper . In fact it also shows the computational limit of them. In particular one can observe that values of the two estimators $\widehat{n}_1$ and $\widehat{n}_2$ of $n$ (respectively based on $D_{n,k}$ and $\widehat{\psi}_{n,k}$) may differ. This should arise especially when most of species are rare and when there were only few abundant species. However, over the fourteen sets of real data we used, this situation occurs four times. Estimations for the three data about tropical beetles seem to be exceptionally satisfactory. It may be due to the presence of many abundant species.

%%%%%%%%%%%%%%%%%%%%%%%%%%%%%%%%%%%%%%%%%%%%%%%%%%%%%%%%%%%%%%%%%%%
% Remerciements
%%%%%%%%%%%%%%%%%%%%%%%%%%%%%%%%%%%%%%%%%%%%%%%%%%%%%%%%%%%%%%%%%%%

%\vspace{0.5cm}

%\paragraph{Acknowledgments} We wish to thank the anonymous referee for encouraging us to improve the paper.

%%%%%%%%%%%%%%%%%%%%%%%%%%%%%%%%%%%%%%%%%%%%%%%%%%%%%%%%%%%%%%%%%%%%%%%%%%%%%%%
% BIBLIOGRAPHIE %%%%%%%%%%%%%%%%%%%%%%%%%%%%%%%%%%%%%%%%%%%%%%%%%%%%%%%%%%%%%%%
%%%%%%%%%%%%%%%%%%%%%%%%%%%%%%%%%%%%%%%%%%%%%%%%%%%%%%%%%%%%%%%%%%%%%%%%%%%%%%%

\nocite{*}

\bibliographystyle{natbib}

\bibliography{species}

%%%%%%%%%%%%%%%%%%%%%%%%%%%%%%%%%%%%%%%%%%%%%%%%%%%%%%%%%%%%%%%%%%%%%%%%%%%%%%%
\end{document}